\documentclass[journal=ancac3,manuscript=article]{achemso}

\usepackage[version=3]{mhchem}
\usepackage[dvipsnames]{xcolor}
\usepackage{soul}
\usepackage{amssymb}
\usepackage{mathrsfs}

\author{Francesco Brasili}
\affiliation{Institute for Complex Systems, National Research Council, Piazzale Aldo Moro 5, 00185, Rome, Italy}
\alsoaffiliation{Department of Physics, Sapienza University of Rome, Piazzale Aldo Moro 5, 00185, Rome, Italy}
\author{Giovanni Del Monte}
\affiliation{Institute for Complex Systems, National Research Council, Piazzale Aldo Moro 5, 00185, Rome, Italy}
\alsoaffiliation{Department of Physics, Sapienza University of Rome, Piazzale Aldo Moro 5, 00185, Rome, Italy}
\author{Angela Capocefalo}
\affiliation{Institute for Complex Systems, National Research Council, Piazzale Aldo Moro 5, 00185, Rome, Italy}
\alsoaffiliation{Department of Physical and Chemical Sciences, University of L’Aquila, Via Vetoio, Coppito, 67100, L’Aquila, Italy}
\author{Edouard Chauveau}
\affiliation{Laboratoire Charles Coulomb, UMR 5221, CNRS–Universit\'{e} de Montpellier, 34095, Montpellier, France}
\author{Elena Buratti}
\affiliation{Institute for Complex Systems, National Research Council, Piazzale Aldo Moro 5, 00185, Rome, Italy}
\alsoaffiliation{Department of Chemical, Pharmaceutical and Agricultural Sciences, University of Ferrara, Via Luigi Borsari 46, 44121, Ferrara, Italy}
\author{Stefano Casciardi}
\affiliation{National Institute for Insurance Against Accidents at Work (INAIL Research), Department of Occupational and Environmental Medicine, Epidemiology and Hygiene, Via di Fontana Candida 1, 00078, Monte Porzio Catone, Rome, Italy}
\author{Domenico Truzzolillo}
\affiliation{Laboratoire Charles Coulomb, UMR 5221, CNRS–Universit\'{e} de Montpellier, 34095, Montpellier, France}
\author{Simona Sennato}
\affiliation{Institute for Complex Systems, National Research Council, Piazzale Aldo Moro 5, 00185, Rome, Italy}
\alsoaffiliation{Department of Physics, Sapienza University of Rome, Piazzale Aldo Moro 5, 00185, Rome, Italy}
\email{simona.sennato@cnr.it}
\author{Emanuela Zaccarelli}
\email{emanuela.zaccarelli@cnr.it}
\affiliation{Institute for Complex Systems, National Research Council, Piazzale Aldo Moro 5, 00185, Rome, Italy}
\alsoaffiliation{Department of Physics, Sapienza University of Rome, Piazzale Aldo Moro 5, 00185, Rome, Italy}

\title{Toward a unified description of the electrostatic assembly of microgels and nanoparticles}


\begin{document}

\begin{tocentry}
\includegraphics{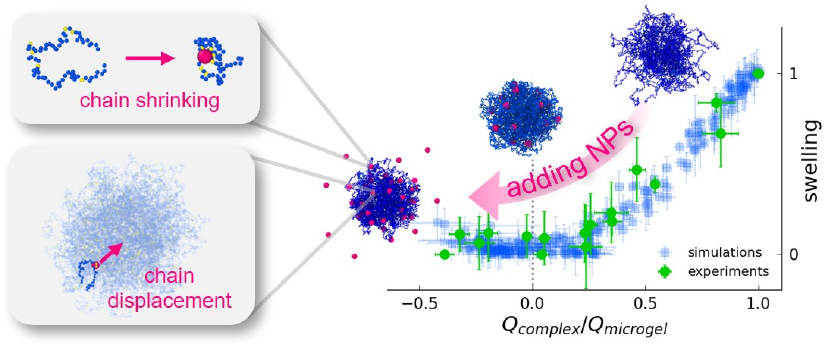}
\end{tocentry}

\begin{abstract}
The combination of soft responsive particles, such as microgels,  with nanoparticles (NPs) yields highly versatile complexes of great potential for applications, from \latin{ad-hoc} plasmonic sensors to controlled protocols for loading and release.
However, the assembly process between these microscale networks and the co-dispersed nano-objects has not been investigated so far at the microscopic level, preempting the possibility of designing such hybrid complexes \latin{a priori}. In this work, we combine state-of-the-art numerical simulations with experiments, to elucidate the fundamental mechanisms taking place when microgels--NPs assembly is controlled by electrostatic interactions. We find a general behavior where, by increasing the number of interacting NPs, the microgel deswells up to a minimum size, after which a plateau behavior occurs. This occurs either when NPs are mainly adsorbed to the microgel corona via the folding of the more external chains, or when NPs penetrate inside the microgel, thereby inducing a collective reorganization of the polymer network. 
By varying microgel properties, such as fraction of crosslinkers or charge, as well as NPs size and charge, we further show that the microgel deswelling curves can be rescaled onto a single master curve, for both experiments and simulations, demonstrating that the process is entirely controlled by the charge of the whole microgel--NPs complex. Our results thus have a direct relevance in fundamental materials science and offer novel tools to tailor the nanofabrication of hybrid devices of technological interest.
\end{abstract}

\section{Introduction}

The design of smart hybrid materials based on the conjugation of stimuli-responsive microgels with nanoparticles (NPs) has attracted increasing interest in recent years, due to the peculiar properties of the two components and to the intriguing features arising from their interplay \cite{ballauff2007,karg2009, arif2021gold}.
Responsive microgels are colloidal particles, made of crosslinked polymeric networks, that are able to change size upon variation of external stimuli such as temperature or pH~\cite{fernandez2011,lyon2012,lee2017}, undergoing a so-called Volume Phase Transition (VPT).
Together with their intrinsic softness, such a responsiveness prompted a wide use of microgels in the development of customized platforms for cutting-edge applications, which range from encapsulation and scaffolding to sensing and pollutant removal \cite{seiffert2014,karg2019,wang2019,naseem2019}. 
Among these, we focus on their potentialities for designing novel hybrid materials where NPs are added as functional building blocks and the polymeric network acts as a stimuli-responsive scaffold.
Of particular interest are microgel--NPs systems assembled using metallic NPs, that provide surface reactivity, catalytic ability and peculiar optical properties. In fact, their resonant absorption of light, occurring at the Localized Surface Plasmon Resonance (LSPR), is strongly sensitive to the dielectric properties at the metal interface and to NPs aggregation~\cite{kelly2003,amendola2017,brasili2020,rossner2021}, enabling the design of photonic nanodevices for biosensing and nanomedicine \cite{willets2007,yu2019,mosquera2020,capocefalo2022}.
Specifically, the combination of plasmonic NPs with thermoresponsive microgels allows for the exploitation of the VPT to trigger the plasmon coupling of NPs in response to specific external stimuli and therefore to gain a strict control on the optical properties of the whole system  \cite{gawlitza2013,manikas2015}. 
The high versatility of such microgel–NPs complexes makes them promising for extremely different applications, including colorimetric temperature sensing\cite{choe2018}, nanoreactors for catalysis\cite{lu2011,wu2012,kakar2021}, photothermal drug-release\cite{cazares2017} and patterning of photonic colloidal crystals\cite{jones2003}.

Despite the wide employment of microgel--NPs systems in applied research, the microscopic interactions between the two components and the changes in their properties upon complexation have not been thoroughly studied. Instead, the fine-tuning of the complexation of such systems would represent a benchmark towards the rational design and engineering of novel devices with optimized properties. 
This gap has not yet been filled, mainly due to the lack of appropriate computer simulation studies.

Recently, some of us put forward a realistic model of microgels~\cite{gnan2017,ninarello2019,camerin2018}, incorporating the disordered and entangled nature of the real particles. Exploiting such a model, in this work we investigate the complexation of microgels with NPs, specifically focusing on the description of the deswelling phenomenon of the polymeric network that is observed upon the addition of NPs.
As a consequence, the microgel--NPs complexes undergo a significant size reduction that could impact the desired functionality.
Such a phenomenon has been recognized in a wide variety of experimental works~\cite{agrawal2013,hou2014,pich2005,hain2008,bradley2011,das2007,davies2010,sennato2021,gawlitza2013}, where NPs of different material, size, shape or surface chemistry, and microgels synthesized using different polymers or co-polymers have been used. Notably, the microgel deswelling is found to occur irrespectively of the method by which NPs are incorporated in the network, including the direct synthesis of NPs inside the microgels~\cite{pich2005,agrawal2013,hou2014,hain2008} and the electrostatic adsorption of charged NPs to oppositely-charged microgels~\cite{das2007,bradley2011,davies2010,sennato2021}. In addition, the microgel deswelling was also found in Ref.~\citenum{gawlitza2013}, where anionic gold NPs were entrapped into anionic poly(N-isopropylacrylamide) (pNIPAM) microgels, by means of the strong, short-range interaction between gold and the amine groups of NIPAM monomers \cite{reimers2016} as driving mechanism. Hence, the phenomenon seems to be independent of the fact that the NPs are located inside the network or mostly attached to the surface~\cite{gawlitza2013}. 

Such a generic finding certainly deserves a convincing microscopic explanation, but so far only unverified hypotheses have been put forward. Most of these rely on some type of favored interactions -- either hydrophobic~\cite{pich2005}, electrostatic~\cite{bradley2011,hain2008,das2007} or specific~\cite{gawlitza2013,hou2014} -- between polymer and NPs, which make the latter to act as additional crosslinkers in the network inducing conformational changes within the microgel. Another explanation, proposed in the literature as either alternative~\cite{agrawal2013} or complementary~\cite{hou2014}, is based on the reduced dynamics of polymer chains due to strong interactions with NPs.
In the case of electrostatic interactions, it has also been suggested that, analogously to an increase in the ionic strength or to a variation of pH~\cite{das2007,davies2010,sennato2021,hain2008}, the presence of NPs could partially shield the charge of ionic monomers, weakening the repulsion between like-charged monomers of the microgel corona that tend to maximize their distance and therefore keep the network stretched. How can all these interpretations explain the fact that such a deswelling is so generic, independently of the employed parameters and, crucially, of the fact that NPs are small enough to diffuse within the network or large enough that they cannot enter and preferentially attach to the surface?

This question is still unanswered and all tentative hypotheses are not yet supported by a microscopic evidence, able to provide a conclusive explanation of the mechanisms underlying the NPs-induced microgel deswelling.
To fill this gap, here we employ a combined numerical and experimental approach, to tackle the problem of the formation of microgel--NPs complexes due to electrostatic interactions. It is worth noting that early simulations of small ionic networks loaded with oppositely-charged NPs were carried out by Quesada-Perez and coworkers, but they mainly addressed the problem of charge inversion~\cite{ramos2019} rather than the deswelling process. In addition, these works assumed an ordered structure for the microgels and an uniform charge distribution, that cannot provide a realistic description of the system. 
In particular, we recently showed that even in pure pNIPAM microgels, the ionic groups of the initiators used in the chemical synthesis, which are preferentially located on the external microgel surface, remarkably affect the VPT~\cite{del2021,elancheliyan2022}.
We now exploit such detailed modeling of the network to study the interplay of a microgel with oppositely charged NPs. We then vary different parameters, namely the total charge of NPs and of the microgel, to modulate their attraction, the size of NPs and the microgel crosslinker concentration to distinguish the situation where NPs stick to the surface or can freely enter the network. Furthermore, we not only consider the microgel having charges predominantly on the surface, mimicking pNIPAM microgels, but also vary their arrangement into a random distribution, more representative of ionic microgels.

Our simulations confirm the occurrence of NPs-induced microgel shrinking under a wide range of conditions, relevant for applications, and allow us to fully elucidate such process, which we interpret in terms of the modifications of the individual polymer chains composing the network upon interactions with the NPs. Furthermore, we are able to rescale all deswelling curves into a unifying master curve using the effective charge of the microgel--NPs complexes as the control parameter of the process.
Given the generality of the results, this work provides the fundamental basis on how to choose the experimental parameters, such size of NPs, charge or crosslinker concentration, in order to tune the formation of microgel--NPs complexes with desired properties for applications.

\section{Results and discussion}

\subsection{Assembly of microgel-nanoparticles complexes: experiments and numerical simulations}
\begin{figure}[t]
\centering
\includegraphics[width=0.95\textwidth]{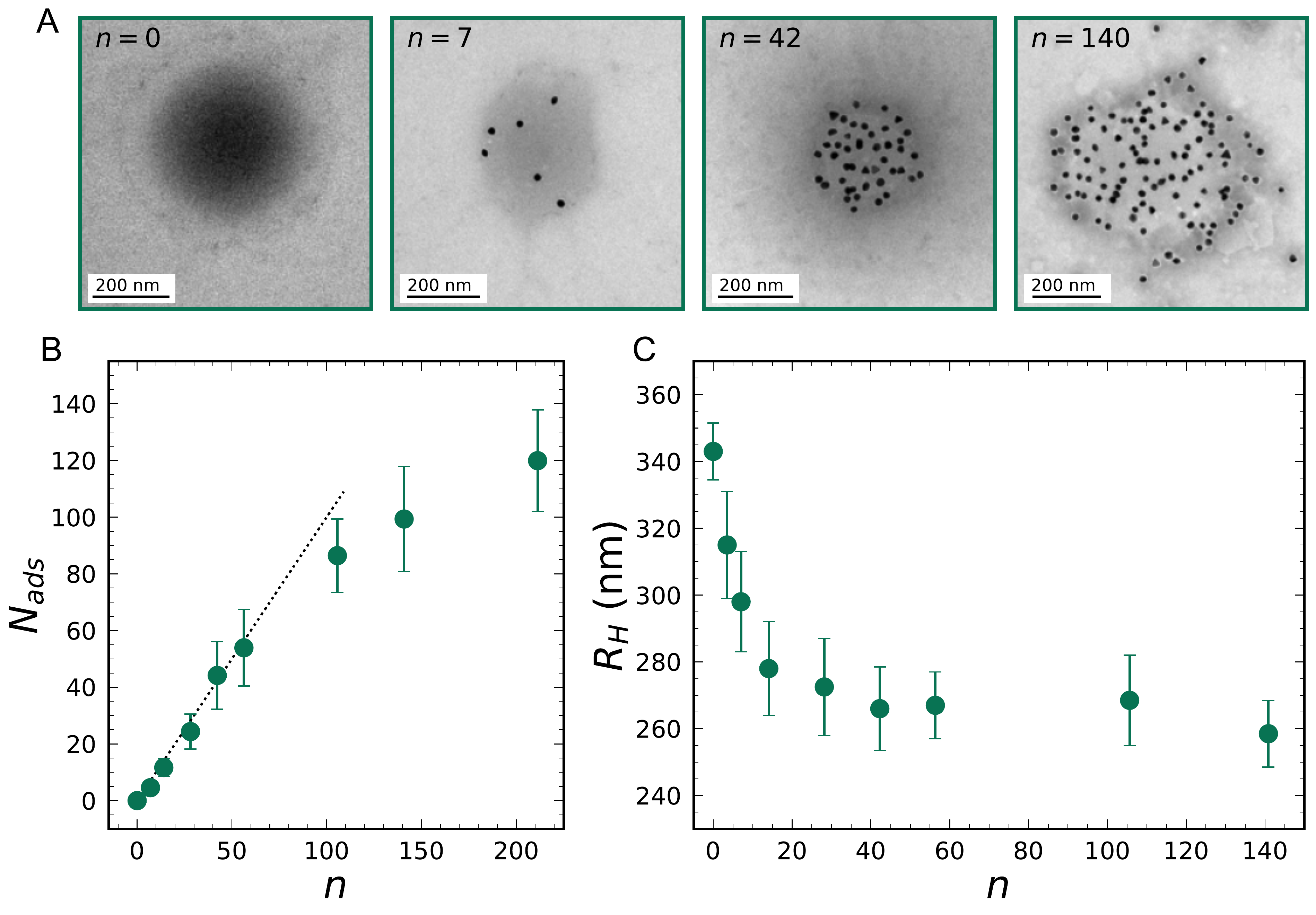}
\caption{Experimental characterization of microgel--NPs complexes, assembled at 25 °C using gold NPs with 20 nm diameter and PNIPAM microgels with fraction of crosslinkers $c = 0.05$: (A) TEM images of microgel--NPs complexes assembled at varying the number ratio $n$ between the two components; (B) Adsorption curve: each value of $N_{ads}$ is calculated from TEM images by counting the number of NPs adsorbed on at least 20 different microgels, error bars are the corresponding standard deviations. The dotted line, $N_{ads}=n$, represents adsorption of all the NPs in the sample; (C) Radial shrinking of microgels induced by adsorption of NPs: the hydrodynamic radius $R_H$ is plotted as a function of $n$.} 
\label{fig:fig1}
\end{figure}

We start by reporting experimental results showing the deswelling of microgels upon addition of oppositely-charged NPs. To this aim, differently from conventional studies, we synthesized pNIPAM microgels using a cationic initiator, as explained in Materials and Methods, so that they easily assemble due to electrostatic interactions with anionic gold NPs. We used molar fractions $f = 0.032$ for charged monomers and $c = 0.05$ for crosslinkers, obtaining microgels with hydrodynamic radius $R_H=343\pm 9$ nm and electrophoretic mobility $\mu_e=0.48\pm 0.04\times10^{-8}$ m$^2$/Vs. We performed dynamic light scattering (DLS) and transmission electron microscopy (TEM) experiments as a function of increasing NPs concentration. The corresponding results are summarized in Figure \ref{fig:fig1} as a function of the NPs--microgels number ratios $n$, ranging from 0 to 200.

 TEM images (Figure \ref{fig:fig1}A) show that most of the NPs (black dots) are homogeneously adsorbed on the microgels (areas more opaque than the background). Only at the highest number ratios some non-adsorbed NPs can be found, as highlighted by the wide field  images shown in the Supporting Information (SI, see Figure S1).
We counted on TEM images the average number $N_{ads}$ of NPs adsorbed to each microgel and plotted it as a function of $n$ in Figure \ref{fig:fig1}B, highlighting that for small $n$ all the dispersed NPs adsorb on the microgels, while $N_{ads}$ deviates from the linear trend for $n\gtrsim 60$, as some NPs remain free in the suspension. 
In this respect, we note that, in correspondence of the same values of $n$, the microgel--NPs complexes undergo charge inversion (Figure S2), giving rise to repulsive interaction with free NPs and therefore reducing their adsorption.
DLS measurements provide the hydrodynamic radius $R_H$ of the microgel--NPs complexes, that is reported in Figure \ref{fig:fig1}C, highlighting the microgel shrinking due to the interaction with the NPs. In particular, a $\sim 20\%$ decrease of $R_H$ is observed upon increasing $n$, from roughly 340 nm down to a plateau value of $\sim 260$ nm, reached for $n \simeq 30$ and maintained afterwards.

The adsorption of NPs and the consequent microgel shrinking in solution are the phenomena on which we focus in this work.
Our first goal is to achieve a coarse-grained representation of the microgel--NPs system that allows us to investigate these phenomena at the microscopic scale. 
We consider microgels with similar characteristics as the experimental ones, thus having a fraction of charged monomers $f=0.032$ and of crosslinkers $c=0.05$. Our coarse-grained modeling, based on a monomer-resolved representation of the polymer network,
accounts for the presence of a disordered network with a realistic core–corona distribution~\cite{ninarello2019,del2019}. In addition, we explicitly include the presence of counterions, providing in this way a proper description of the local inhomogeneities arising in the network due to their reversible binding~\cite{del2019}. 
With this model we perform simulations of a single microgel with $N\sim 14000$  monomers of diameter $\sigma$, whose total size can be quantified by its hydrodynamic radius $R_H$, that is calculated in simulations using the ZENO algorithm~\cite{zeno}. The microgel interacts with a variable number $n$ of NPs with diameter $D=2\,\sigma$, to achieve the same size ratio between microgel and NP as in experiments, and charge $q=-35\,e$, to match the experimentally determined value (see SI, Figure S3). The microgels, mimicking  pNIPAM ones where the charged groups solely come from initiators, have charges only on the surface~\cite{del2021}. We work in good solvent conditions, so that monomers interact with the standard bead-spring model of Kremer and Grest~\cite{kremer1990}. For the interactions between NPs and the monomers of the microgel, we adopt an excluded-volume repulsion plus the electrostatic contribution, which in our oppositely-charged system is the dominant interaction, thus neglecting other effects, such as hydrophobic or specific ones. A detailed description of the model, interaction potential, simulation units and employed parameters is given in Materials and Methods.
\begin{figure}[htbp]
\centering
\includegraphics[width=0.95\textwidth]{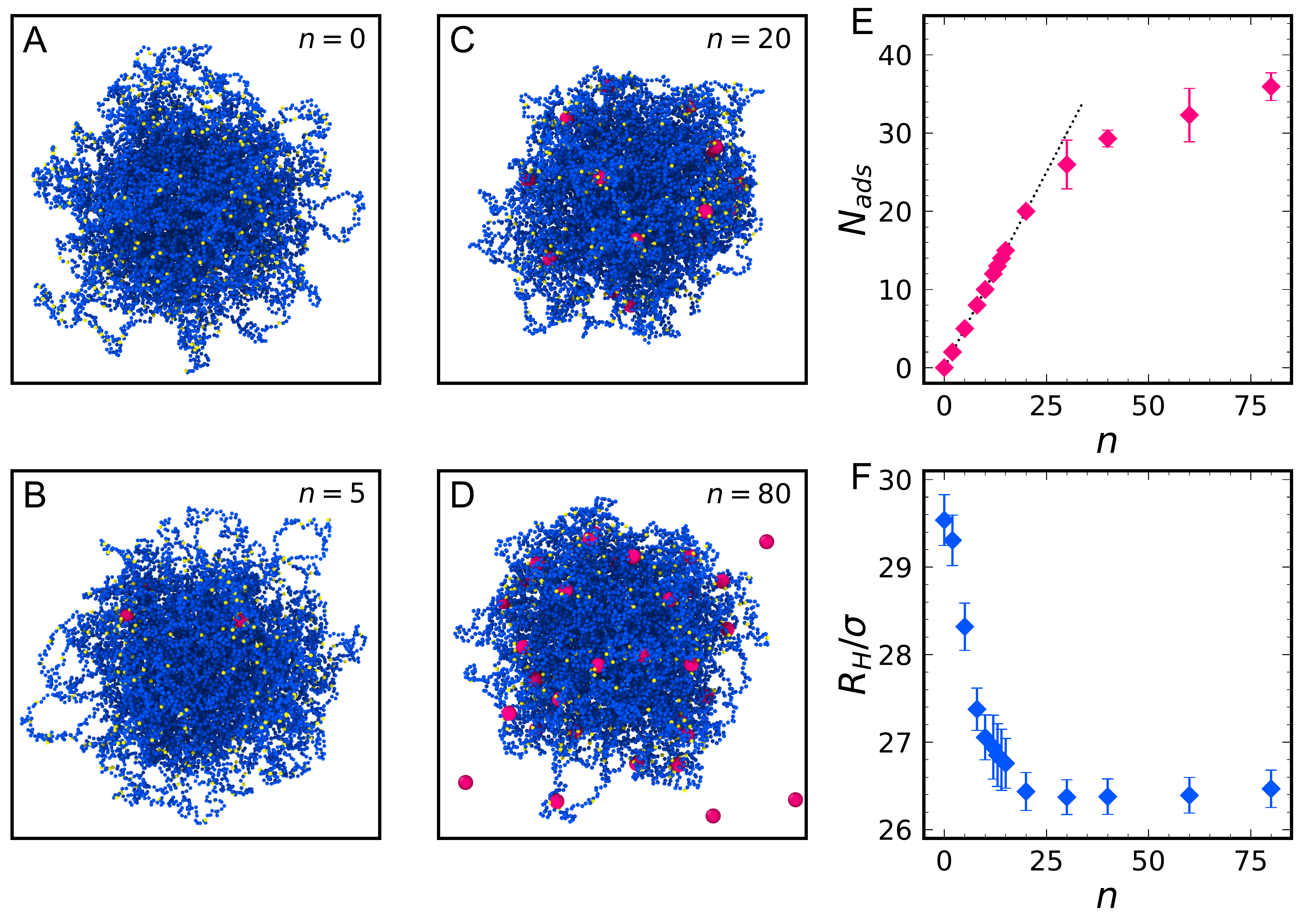}
\caption{Molecular dynamics simulation results for microgel--NPs systems varying the number $n$ of NPs, corresponding to the experimental results of Figure \ref{fig:fig1}. The microgels are synthesized \latin{in silico} with $c=0.05$ and $f=0.032$, indicating crosslinker and charge fraction, respectively; NPs have diameter $D=2\,\sigma$ and charge $q=-35\,e$: (A-D) simulation snapshots at selected $n$ values; blue, yellow and red particles represent neutral monomers, charged monomers and NPs, respectively; (E) adsorption curve: each value of $N_{ads}$ is calculated selecting the NPs whose distance from the microgel center of mass is lower than $R_H$, and averaging their number over $10^3$ configurations, within a simulation time of $2\times10^7\tau$. Error bars are the corresponding standard deviations; the dotted line, $N_{ads}=n$, represents adsorption of all NPs; (F) radial shrinking of microgel--NPs: the hydrodynamic radius $R_H$ of the microgel--NPs complex, calculated through the ZENO algorithm, is plotted as a function of $n$. }
\label{fig:fig2}
\end{figure}

Numerical results, corresponding to the experimental ones in Figure \ref{fig:fig1}, are summarized in Figure \ref{fig:fig2}. From the snapshots, it can be seen that the polymer chains protruding out of the dense core of the microgel in absence of NPs progressively become more and more incorporated within the main network, when NPs are added in the simulation box.
At the highest studied values of $n$, in agreement with electron microscopy, it is also evident that some NPs are not embedded in the polymer matrix and instead freely diffuse within the available volume.
To derive the adsorption curve, we identify adsorbed NPs as those whose distance $r_{\text{NP}}$ from the center of mass of the microgel satisfies the condition $r_{\text{NP}}<R_H$.
The number $N_{ads}$ of adsorbed NPs is plotted as a function of $n$ in Figure \ref{fig:fig2}E.
All NPs adsorb to the microgel at low $n$, while $N_{ads}<n$ for $n\geq30$, similarly to what observed in experiments.
The NPs-induced shrinking of the microgel is demonstrated by the behavior of the hydrodynamic radius $R_H$ of the microgel--NPs complexes, that is reported as a function of $n$ in Figure \ref{fig:fig2}F. $R_H$ shows a decreasing trend down to a plateau value that is reached for $n\simeq 20$. The qualitative agreement of these results with experiments validates the modeling of the microgel--NPs systems adopted for simulations and highlights the role of electrostatic interactions, that are sufficient to reproduce NPs adsorption and related microgel deswelling.

\subsection{Radial shrinking: incorporation of NPs and microscopic interpretation}
\begin{figure}[t!]
\centering
\includegraphics[width=0.95\textwidth]{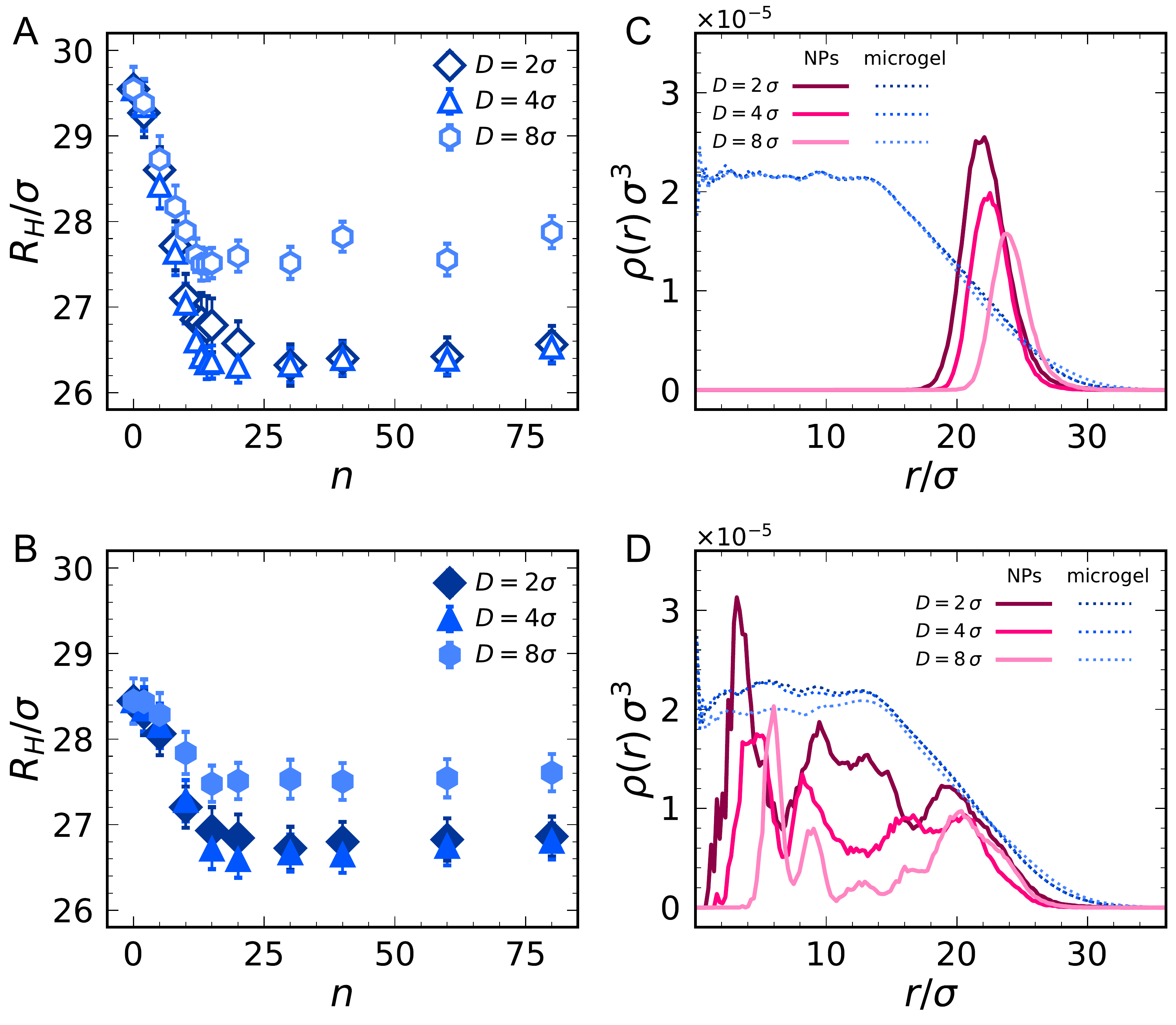}
\caption{Microgel shrinking and penetration of NPs within the microgel: $R_H$ of the NPs–microgel complex is plotted as a function of $n$ for three NPs diameters (as labelled in the figure) for 
microgels with surface charges (A) and with random charges (B); radial profiles of NPs (full lines) superimposed to those of microgel monomers (dotted lines) in the case $n=40$ (colors corresponding to the different NPs diameters) for surface (C) and random (D) charge distributions.  Data are averaged over five different microgel topologies to reduce numerical noise.}
\label{fig:fig3}
\end{figure}

Having established our simulation protocol, we now aim to achieve a detailed comprehension of the mechanisms underlying microgel complexation with NPs. To start with, we focus on the radial shrinking of the microgel that occurs upon addition of NPs with different diameters: $D=2\,\sigma$, $D=4\,\sigma$ and $D=8\,\sigma$ and microgels with different spatial distributions of the charged monomers: surface and random. The behavior of the hydrodynamic radius as a function of $n$ is reported in Figure \ref{fig:fig3}A and Figure \ref{fig:fig3}B for surface and random charge arrangements on the microgel, respectively. In all analyzed cases, a decreasing trend of $R_H$ is observed with increasing $n$, highlighting the generality of the radial shrinking of microgel upon incorporation of NPs. The microgel shrinking is rapid at small $n$, while for sufficiently large $n$ ($n\gtrsim 20$) a plateau is reached in all cases. However, comparing data within the same charge distribution, we find that the plateau of $R_H$ is higher for NPs of larger size.
Instead, the spatial distribution of charged monomers affects both the initial size of the microgel (the value of $R_H$ for $n=0$ is higher in the case of surface charge distribution) and the overall deswelling. Indeed, the relative degree of shrinking for $D=2\,\sigma$ is roughly 10\% for surface microgels and $\sim$ 5\% for random charge distribution, signaling a less efficient mechanism.

To understand these results, it is first important to analyze the location of the NPs within the polymer network for the different studied conditions. This can be clearly visualized by superimposing the radial density profiles $\rho(r)$ of the NPs onto those of the microgel monomers, as shown in Figure \ref{fig:fig3}C and \ref{fig:fig3}D, for surface and random charge distributions, respectively, focusing for simplicity on simulations with $n=40$. Data for other values of $n$ display similar features.
As well-known, the microgel structure is characterized by a uniformly dense core (approximately, for $r \leq 15\,\sigma$) and by an external, soft corona. The spatial arrangement of added NPs is largely different in the two cases, since they tend to locate in the proximity of charged groups: they remain strongly adsorbed on the microgel corona for surface distribution, while they are able to penetrate within the microgel core for random charges. In both cases, the NPs size affects the depth of penetration within the microgel, since smaller NPs can reach inner regions of the microgel where the polymer network is tighter. This is evident in the density profiles of the NPs, showing a Gaussian-like arrangement within the corona for the surface case and a sequence of oscillations within the whole microgel structure for the random one. The latter behavior clearly indicates the structuring of the NPs within the randomly-charged microgel, which can penetrate more and more as their size decreases. 

The different localization between small ($D=2\,\sigma$ and $D=4\,\sigma$) and large NPs ($D=8\,\sigma$) also explains the different plateau value of $R_H$ reached at high $n$. Indeed, at least part of the large NPs are located in the very external part of the corona, particularly for surface charges, when the tail of the NPs profile exceeds the monomer profile and a significant amount of NPs is found at $r>R_H$. These NPs form an external layer that, at high $n$, directly contributes in hindering the contraction of the microgel corona, thus leading to a re-increase of $R_H$ with respect to the minimum value reached upon full adsorption.

The depth of penetration of NPs within the microgel also explains the larger radial shrinking of the microgel observed for surface-charged microgels than for random ones. In the first case, being the microgel more swollen by itself, its corona is less dense, so that the NPs attaching to it have a much larger effect on its shrinking. Conversely, when they penetrate, in the random case, still there is a decrease of the microgel size, but it is overall less pronounced.  
\begin{figure}[h!]
\centering
\includegraphics[width=0.95\textwidth]{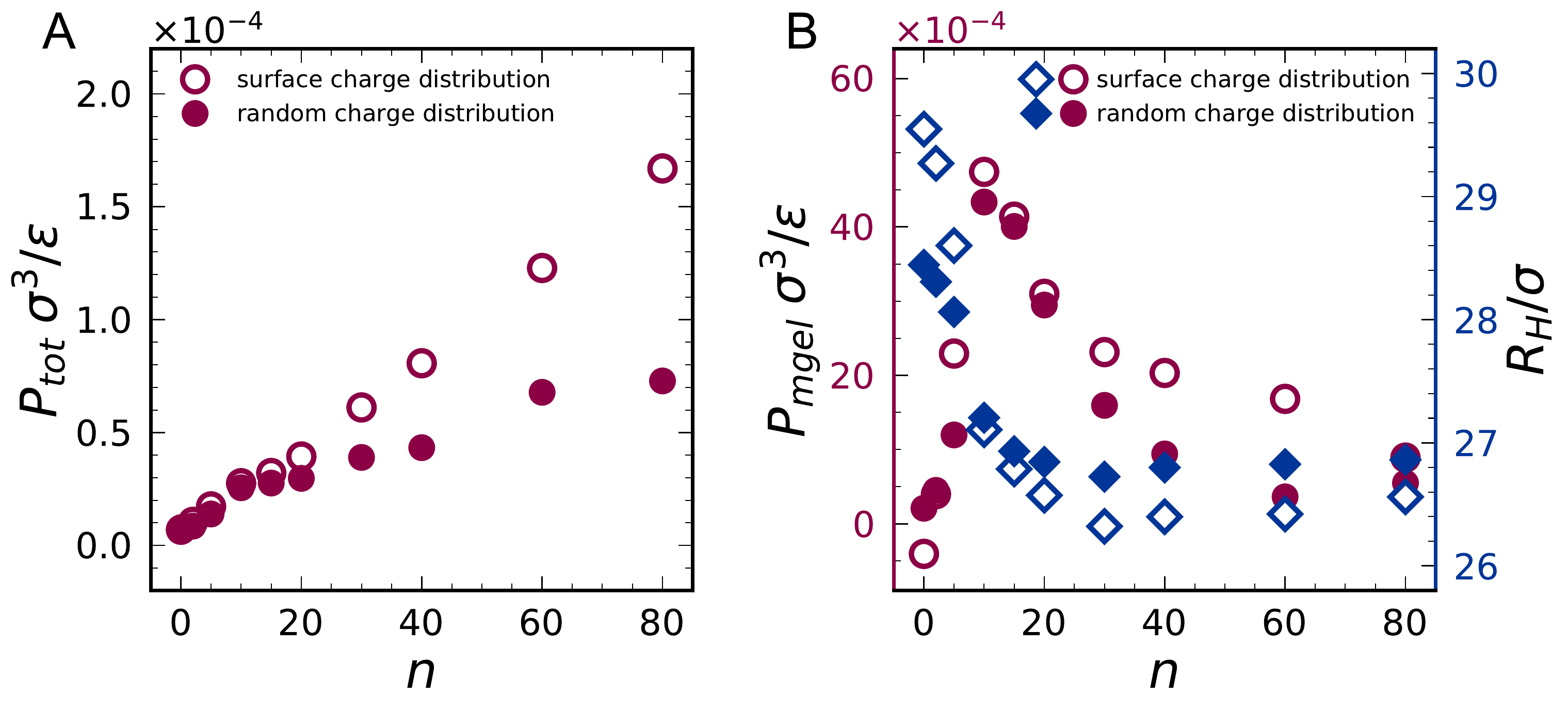}
\caption{Pressure as a function of $n$, in the case of NPs with $D=2\,\sigma$ and $q=-35\,e$. The pressure $P_{tot}$ of the total system (A) and $P_{mgel}$ of the microgel only (B) are shown for the surface (hollow circles) and random (full circles) charge distributions. The corresponding trends of the hydrodynamic radius are also shown in panel B.}
\label{fig:fig4}
\end{figure}

The underlying physical mechanism causing this observation can be understood by looking at the pressure behavior upon addition of NPs. In Figure \ref{fig:fig4}A, we report the total pressure of the system, $P_{tot}$, as a function of $n$. It starts from zero for $n=0$, because the microgel is fully-bonded (hence the negative contribution of bonds exactly counteract the repulsive ones from steric interactions and thermal motion), while the contribution of counterions is much smaller. Then, as expected, $P$ increases monotonically with $n$ in all cases, independently of NPs being or not completely adsorbed, penetrating within the microgel or not.  For surface-charged microgels, we can consider that this extra pressure is mainly due to the NPs, exerting an osmotic-like contribution,  since they are all located outside the microgels, pushing the microgel to shrink. 
For random-charged microgels, this effect is weakened since some NPs are also inside the network, thus the net osmotic contribution is lower. This explains why $P_{tot}$ is smaller at large $n$ for the random case.

\begin{figure}[h!]
\centering
\includegraphics[width=0.95\textwidth]{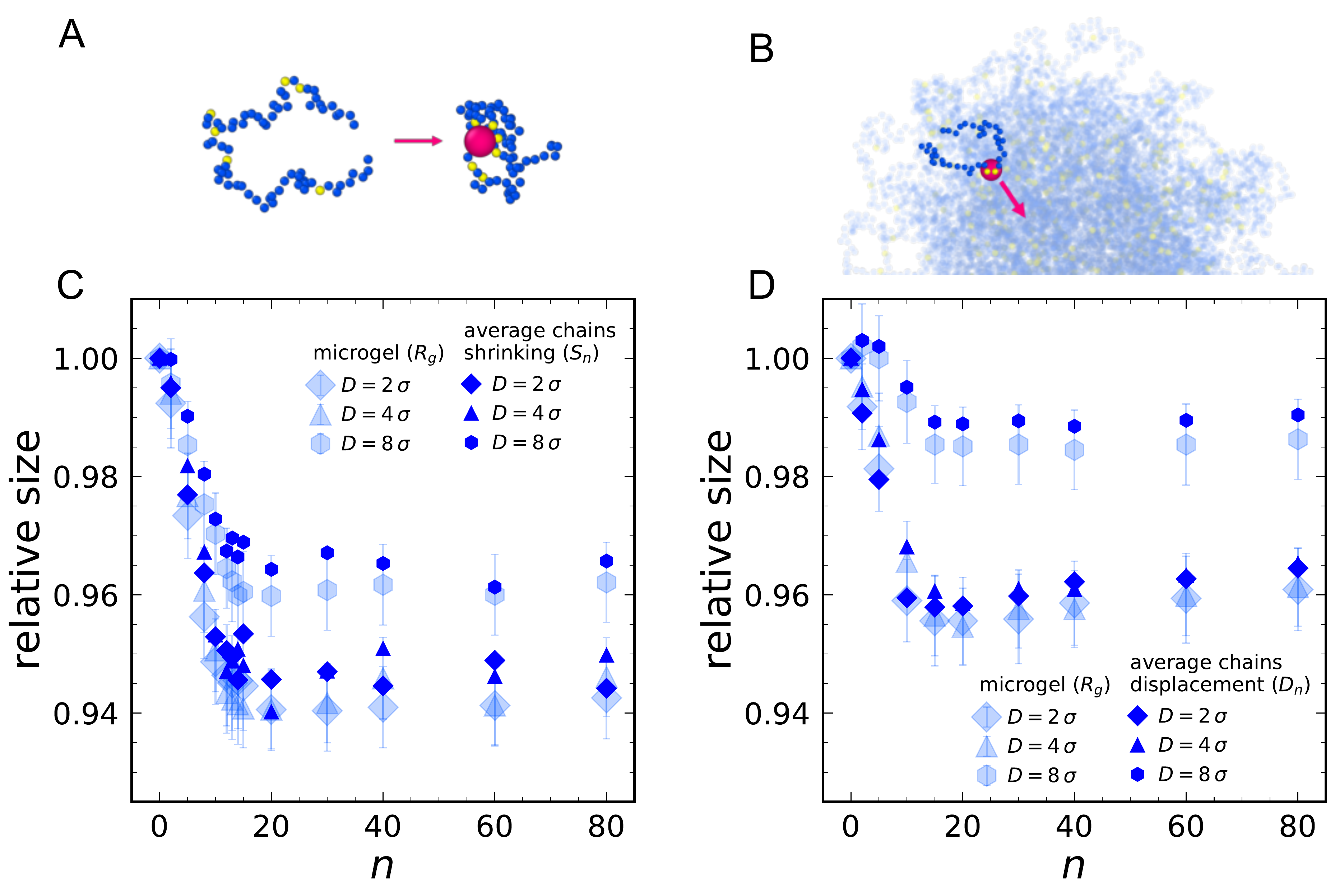}
\caption{Microscopic interpretation of the microgel shrinking. Sketches of the two main mechanisms by which the shrinking occurs: (A) chain shrinking and (B) chain displacement; blue, yellow and red particles represent neutral monomers, charged monomers and NPs, respectively;  Radial shrinking of the microgel $R_g/R_g^{(0)}$ (transparent symbols) as a function of $n$, compared to the microscopic quantities $S_n$ and $D_n$ (defined in Eq.~\ref{eq:microscopic}, full symbols) for surface (C) and random (D) charge distributions, respectively. The trends for three NPs sizes, indicated in the labels, are reported in each plot.}
\label{fig:fig5}
\end{figure}

However, it is most instructive to focus on the pressure acting on the microgel only, $P_{mgel}$, shown in Figure \ref{fig:fig4}B, again as a function of $n$. First, we notice that $P_{mgel}$ starts from a slightly negative value for the surface charge distribution case, whereas it starts from roughly zero for random conditions. This is due to the different organization of negative (microgel) counterions for the two charge distributions, reported for completeness in the SI (Figure S4). Indeed, in the surface case, such counterions mostly remain in the outer part of the microgel, resulting in the stretching of the microgel corona that is, somehow, pulled outwards by the electrostatic attraction.
This also contributes to the higher microgel size with respect to the random case.
In presence of NPs, we then find a non-monotonic behavior of $P_{mgel}$ up to $n\simeq 10$, that roughly corresponds to the end of the rapid radial shrinking, also reported in Figure \ref{fig:fig5}B. The maximum pressure thus occurs when all NPs are directly interacting with the microgel, after which there is an excess of NPs which still interact with long-range Coulomb interactions with the microgel--NPs complex, which have an opposite effect. Indeed, they slightly pull the microgels via their charged monomers, as also visible in the snapshot with $n=80$ reported in Figure \ref{fig:fig2}.
Hence the shrinking tendency is reduced at high enough $n$, because the pressure on the microgel is lower and tends to approach zero again, signaling that the microgel in the complex with NPs tends to reach again a fully relaxed state.

We now investigate the microscopic mechanisms causing the microgel shrinking upon incorporation of NPs, by analyzing the single polymer chains composing the network.
The chains are defined as the sets of consecutively bonded divalent monomers that connect two crosslinkers. Each chain can be described by four quantities: its length $\ell_k$, namely the number of monomers comprising it, its total charge $q_k$, counting the number of charged monomers in the chain, the radial position in the microgel $r_k$, defined as the distance of the chain center of mass from that of the microgel, and finally the gyration radius $R_{gk}$.
The analysis of the chain composition of the microgel, previously reported in Ref.~\citenum{del2021}, clearly indicates that longer and more charged chains are mainly located in the microgel corona. This effect is more pronounced for surface than for random distribution. 

We are now interested in the modification of the chains conformation and positioning within the microgel, expressed by the parameters $R_{gk}$ and $r_k$, and on their influence on the overall shrinking of the microgel. Therefore, we derive, as detailed in the SI (subsection S2.2), an expression for the gyration radius $R_g$ of the whole microgel in terms of the chains parameters $r_k$ and $R_{gk}$:
\begin{equation}
\label{eq:rg_chains}
R_g \simeq \sqrt{\frac{1}{N}\sum_k R_{gk}^2 \ell_k + \frac{1}{N}\sum_k r_k^2 \ell_k} \quad ,
\end{equation}
where the sums are computed over all the chains of the microgel, and the contribution of crosslinking monomers, non-zero only in the second term, is neglected. This is a suitable approximation as shown in Figure S5. 

Equation~\ref{eq:rg_chains} discloses the two main contributions to the size of the whole microgel.
A shrinking of the microgel can therefore arise either from a contraction of the chains or from their displacement towards the center of mass of the microgel. These two mechanisms are schematically illustrated in Figs.~\ref{fig:fig5}A and \ref{fig:fig5}B.

To better distinguish the two contributions and their weights in the NPs induced shrinking of the microgel, we define two simplified quantities for the microgel interacting with $n$ NPs, the average chains shrinking $S_n$ and the average chains displacement $D_n$, as
\begin{equation}
\label{eq:microscopic}
S_n = \frac{ \left< \displaystyle\sum_k \ell_k\, R_{gk}^{(n)} \right> }{ \left< \displaystyle\sum_k \ell_k\, R_{gk}^{(0)} \right> }
\qquad \text{and} \qquad
D_n = \frac{ \left< \displaystyle\sum_k \ell_k\,    r_k^{(n)} \right> }{ \left< \displaystyle\sum_k \ell_k\,    r_k^{(0)} \right> }\quad ,
\end{equation}
where $n$  refers to the number of NPs in the simulation, thus the quantities in the denominators are computed on the bare microgel, and the brackets $\left<\cdot\right>$ indicate ensemble averages.

In Figure \ref{fig:fig5}C, the evolution of $S_n$ with $n$ is superimposed to the shrinking $R_g^{(n)}/R_g^{(0)}$ of the gyration radius of the microgel for surface charge distribution and all studied values of NP diameter. Instead, for random charges, the microgel shrinking is well-captured by $D_n$, as shown in  Figure \ref{fig:fig5}D. For completeness, we also show the comparison of the microgel shrinking with the other quantities ($D_n$ for surface distribution and $S_n$ for random one) in the SI (Figure S6), showing that they are not able to describe the data. Only for the random case and $D=8\,\sigma$, the microgel shrinking is also well-described by $S_n$. 

Based on these observations, we hypothesize that when NPs are adsorbed only onto the microgel corona (surface charge distribution or large NPs), the driving microscopic mechanism for microgel deswelling is the deformation of individual corona chains that tend to wrap around the NPs.
Instead, when NPs are able to penetrate within the microgel core (random charge distribution and small NPs), they tend to drag chains inwards. This is a collective effect, because of the strong connectivity of the microgel.
In this respect, a full analysis of the modifications induced by incorporation of NPs at the level of single polymer chains reveals that shrinking and displacement are more pronounced for charged chains than for neutral ones, and they decrease for decreasing net charge and length of the chains.
Furthermore, in the case of charged chains, the shrinking is more pronounced when the distance from a NP decreases, while for neutral chains it is independent of  the distance from NPs.
This confirms that the shrinking of chains is the main mechanism responsible for the size reduction of the overall microgel if the NPs interact with chains that are long enough to bend quite freely, without hindrance due to the constraints of crosslinkers, and have large number of charges that are attracted towards the NP surface. Otherwise, the microgel shrinking is due to the reshuffling of the chains that, on average, are pushed closer to the microgel center of mass.

\subsection{Varying parameters: an unifying behavior?}
We now analyze the role of other important parameters in the assembly of microgels and NPs. Namely, we also consider the effect of crosslinker concentration, in order to address microgel topology, and of microgel charge, in order to vary the NP-microgel electrostatic interactions and the resulting assembly.

\begin{figure}[H]
\centering
\includegraphics[width=0.95\textwidth]{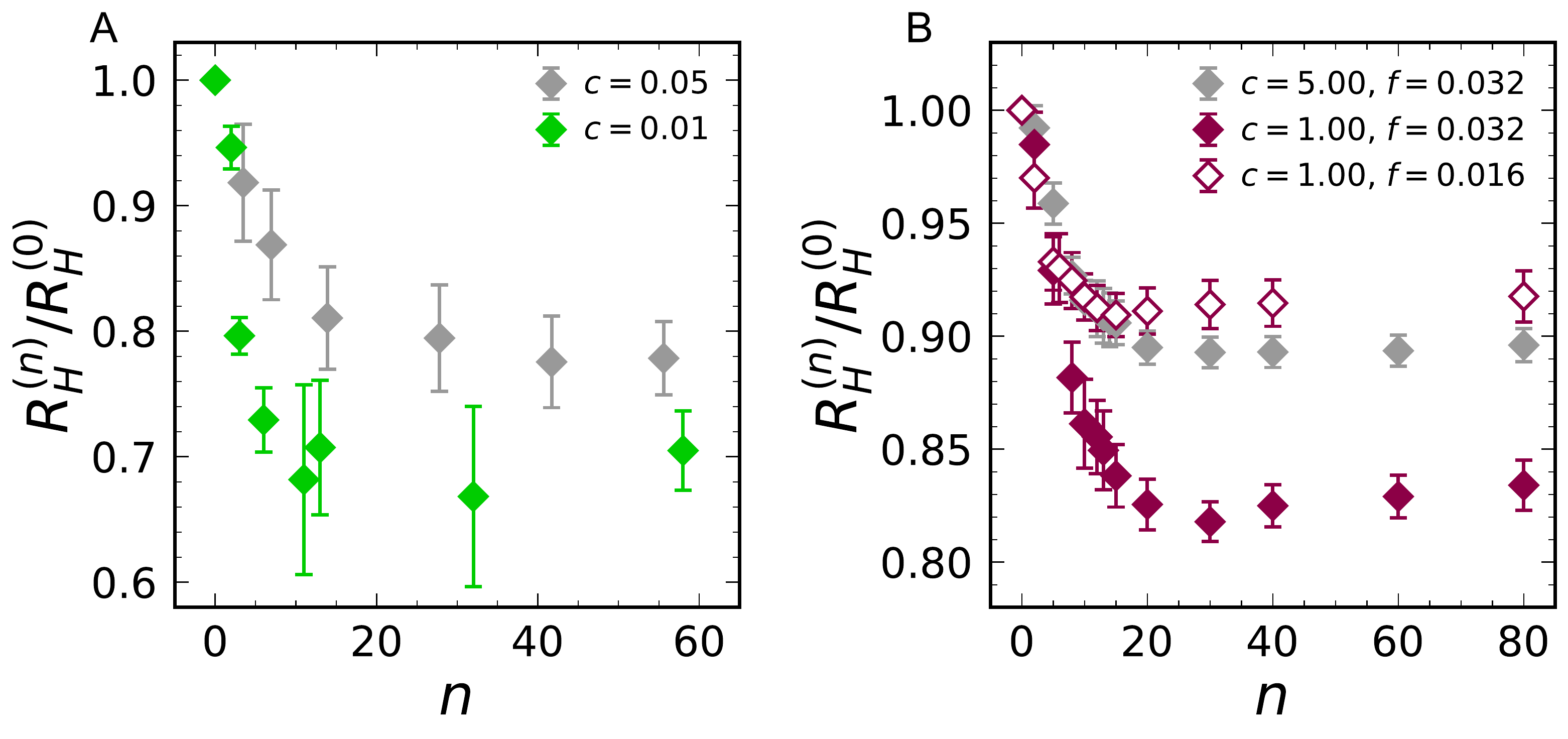}
\caption{Effect of the fraction of crosslinkers on the radial shrinking: experiments (A) and simulations (B). Simulations are performed using surface charge distribution and NPs with $q=-35\,e$ and $D=2\,\sigma$; for $c=0.01$, two fractions of charged monomers, $f=0.032$ and $f=0.016$, are analyzed.}
\label{fig:fig6}
\end{figure}

To this aim, we also synthesized cationic microgels with $c=0.01$ and $f=0.02$, as described in Materials and Methods, and we repeated the experiments in the presence of the same NPs. The resulting hydrodynamic radius is reported in Figure \ref{fig:fig6}A as a function of $n$, again indicating a similar shrinking behavior as observed for $c=0.05$.
From the considerations expressed above, the variations in the two parameters introduce two competing contributions.
The lower degree of crosslinking results in a microgel corona made of much longer chains, giving rise to a more pronounced shrinking, as discussed in the previous paragraph.
On the other hand, the lower fraction of charged monomers reduces the interaction of polymer chains with the adsorbed NPs. We then perform additional simulations for low-crosslinked microgels ($c=0.01$) with two different charge contents,  $f=0.032$  and $f=0.016$, always for the case of surface charge distribution, that more closely resembles the experimental case. We find confirmation that when decreasing $c$ but keeping the charge amount on the microgel the same, the shrinking is much more enhanced, as shown in Figure \ref{fig:fig6}B, where the $f=0.032$ results are plotted for both studied values of crosslinker concentration. However, when decreasing the microgel charge by half, the two effects compensate and thus the deswelling is very much reduced,  going back to a quite similar behavior for both microgels, in good qualitative agreement with the experimental observations.
Of course, a quantitative comparison is not possible since the microgels synthesized in silico are much smaller than the real ones, so that their shrinking is always much less pronounced than in experiments. 
Nonetheless, these results help us to interpret the observations and suggest possible ways to be able to predict the amount of shrinking by a judicious choice of the synthesis parameters.

\begin{figure}[H]
\centering
\includegraphics[width=0.95\textwidth]{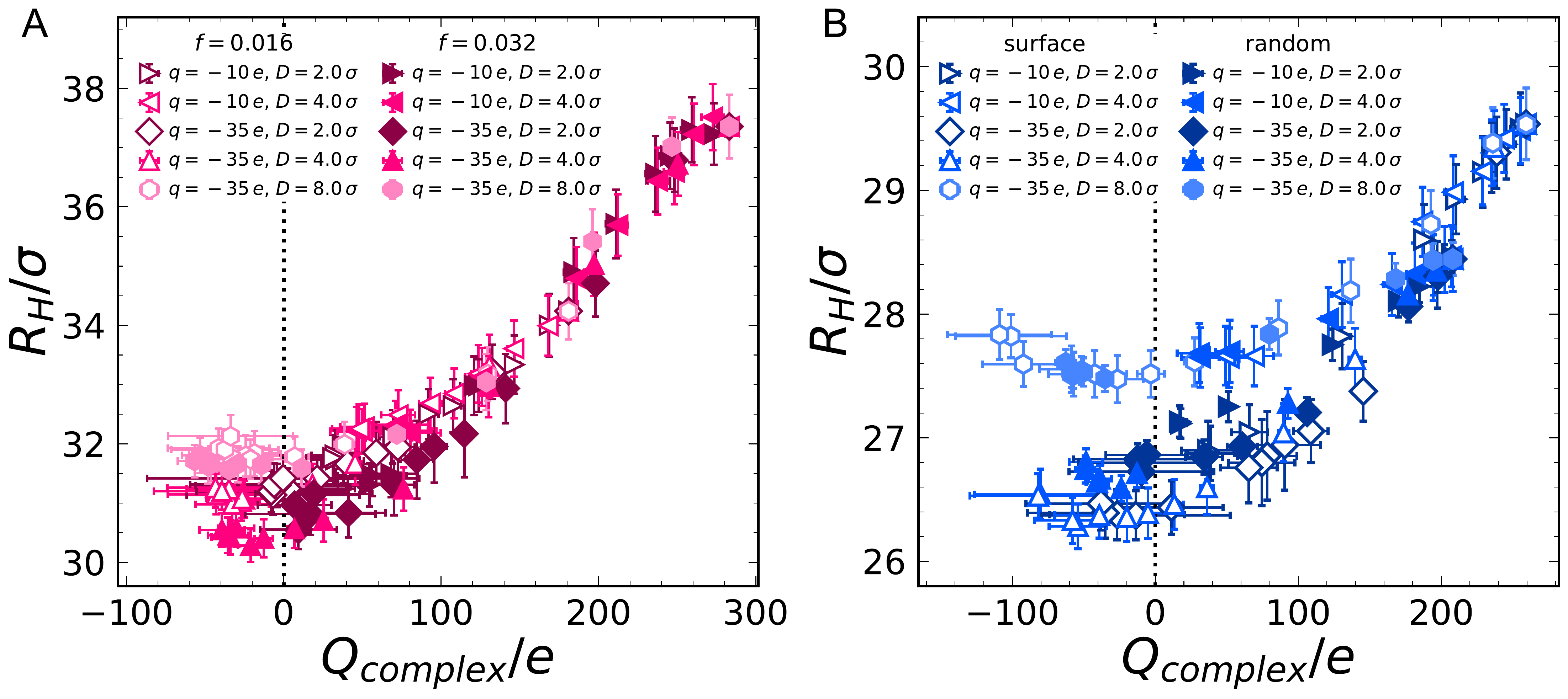}
\caption{Shrinking curves as a function of the total charge $Q_{complex}$ of the microgel--NPs complex. The hydrodynamic radius $R_H$ is plotted for microgel with $c=0.01$ (A, two fractions of charged monomers) and $c=0.05$ (B, two spatial distributions of charged monomers), using NPs with varying charge ($q=-10\,e$ and $q=-35\,e$), and size ($D=2\,\sigma$, $D=4\,\sigma$ and $D=8\,\sigma$).}
\label{fig:fig7}
\end{figure}

To this aim, it is important to identify the right physical observable that would be able to describe the results in a more unifying way. This turns out to be the charge of the microgel--NPs complex $Q_{complex}$, that is defined considering those NPs and counterions attached to the microgel as discussed in Materials and Methods. To show that this is the case, we perform additional simulations of microgels with surface charge distribution, varying the microgel charge $f$, the NPs charge $q$ and the NPs size $D$. The resulting $R_H$ for all simulated microgels are plotted as a function of $Q_{complex}$ in Figure \ref{fig:fig7}A, rescaling onto a single master curve. This confirms that it is the effective charge of the complex the main parameter determining the response of the microgel to the addition of NPs that remain attached to the surface. Furthermore, the plot strongly suggests that the plateau in $R_H$ is reached when $Q_{complex}\approx 0$, suggesting that once neutrality of the complex is attained, further addition of the NPs does not affect the microgel size any longer and, since the charge of the microgel--NPs complex becomes negative, most NPs just go in suspension.

\begin{figure}[H]
\centering
\includegraphics[width=0.95\textwidth]{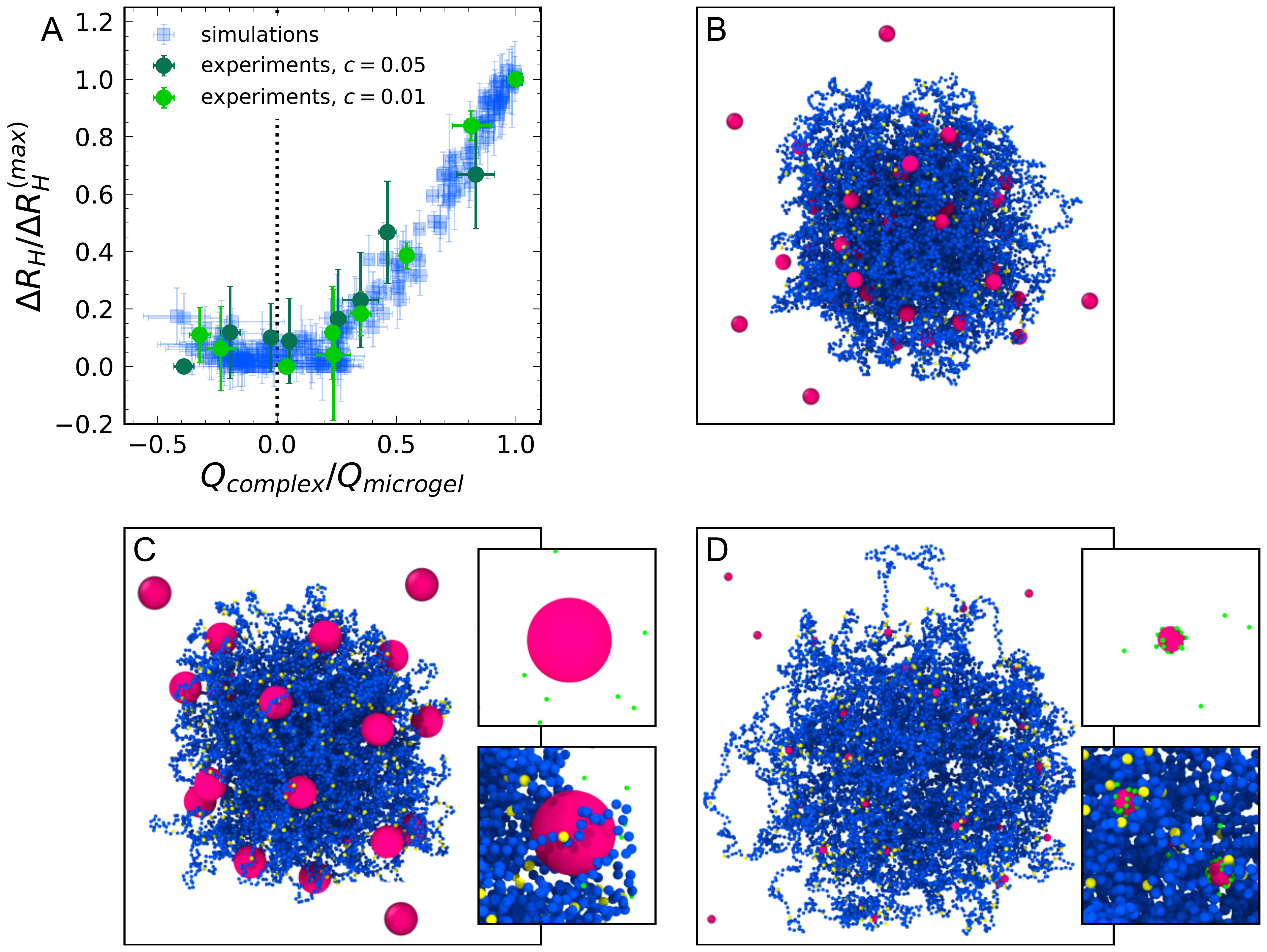}
\caption{(A) Master plot of the shrinking behaviour of the microgel--NPs, obtained by plotting the variation $\Delta R_H/\Delta R_H^{(max)}$ as a function of $Q_{complex}/Q_{microgel}$ for all the types of microgels and NPs simulated in this work and corresponding experimental trends, derived from measurements of $R_H$ and $\mu_e$; selected snapshots of microgel--NPs complexes in different scenarios: microgel with $c=0.05$ and random charge distribution, NPs with $q=-10\,e$ and $D=4\,\sigma$ (B); microgel with $c=0.05$ and surface charge distribution, NPs with $q=-35\,e$ and $D=8\,\sigma$ (C); microgel with $c=0.01$ and $f$, NPs with $q=-35\,e$ and $D=2\,\sigma$ (D). In all cases, blue, yellow and red particles represent neutral monomers, charged monomers and NPs, respectively. All the microgels have a fraction of charged monomers $f=0.032$, interact with $n=80$ NPs, and are shown with the same magnification. The insets of panels C and D show zoomed images of NPs, highlighting the different coverage by positive counterions (green particles) depending on $D$. }
\label{fig:fig8}
\end{figure}

It is now legitimate to ask whether this description also applies to the case where NPs can penetrate within the network. To this aim, we report in Figure \ref{fig:fig7}B the evolution of $R_H$ with $Q_{complex}$ for microgels with the same amount of charge and crosslinker concentration ($f=0.032$ and $c=0.05$), but now comparing surface and random charge distributions.
In this case, the two behaviours differ in the overall range, and more evidently at high $n$.
The lower effective charge $Q_{complex}$ of the complexes in the case of random distribution is due to the more efficient internalization of the negative (NPs) counterions (Figure S4). This is also reflected in the lower microgel size and in the lower values of $P_{tot}$, as previously discussed.
In the limit of high number of attached NPs, the trends reach different plateau values of $R_H$.
Specifically, for small NPs ($D=2\,\sigma$ and $D=4\,\sigma$), the final size of the microgels differs between charge distributions, while, when $D=8\,\sigma$, a common, markedly larger value is reached.
These differences correspond to the different internalization behavior of the adsorbed NPs (penetration in the core for random distribution and small NPs, external attachment for surface distribution or large NPs), discussed previously.
Nevertheless, a common feature of all the studied systems can be recognized noting that the minimum of $R_H$ is reached for $Q_{complex}\approx 0$. Therefore, in order to incorporate these dependencies into a unifying picture, we define the normalized swelling amount as $\Delta R_H/\Delta R_H^{(max)}=(R_H-R_H^{(min)})/(R_H^{(max)}-R_H^{(min)})$, where $R_H^{(max)}$ and $R_H^{(min)}$ are the maximum and minimum values of $R_H$, and
plot it as a function of $Q_{complex}/Q_{microgel}$ in Figure \ref{fig:fig8}A.
Here, we also include $c=0.01$, whose NPs-free limit has a different value of $R_H$, and obtain a final unifying plot. 
We find that, within the numerical error, a single master curve, represented by transparent blue markers, is observed. We also note that a similar result holds for $R_g$ (see Figure S7), confirming the robustness of our findings.
Experimentally, we can build the same curve by considering the measured $R_H$ of the microgels and the effective electrokinetic charge ratio $Q^{(exp)}_{complex}/Q^{(exp)}_{microgel}$, derived from the measurements of $\mu_e$ (Figure S2) using Henry's formula, given by Eq.~\ref{eq:henry}.
The experimental trends obtained for the two fractions of crosslinker are superimposed to the numerical master curve in Figure \ref{fig:fig8}A, showing an excellent agreement. This corroborates our results in silico and confirms that $Q_{complex}$ is the main quantity determining how much a microgel shrinks in the presence of NPs with respect to its NP-free state, independently of both the microgels and the NPs specific characteristics.
Further, the remarkable agreement between experiments and simulations points to a novel predictive tool for the amount of adsorbed particles on charged microgels based only on measurement of electrophoretic mobility, being $N_{ads}=n$ when $Q_{complex}/Q_{microgel}>0$. On the other hand, it designates the isoelectric condition as the one delimiting the onset of NPs excess in the bulk. This turns out to be decisive when the adsorbing efficiency of microgels has to be taken into consideration.

Such a result is quite remarkable if one thinks that the present investigation covers very different scenarios, for which some examples are reported in the representative snapshots of Figure \ref{fig:fig8}B-D, all corresponding to $n=80$. In particular, Figure \ref{fig:fig8}B displays a microgel with random charge distribution and intermediate NP size $D=4\,\sigma$, where it can be seen the incorporation of NPs inside the network and several NPs free in suspension. The fact that the NPs are not completely localized within the corona, makes the latter rather fluffy and disordered, even in saturated conditions. On the other hand, in the second selected situation, reported in Figure \ref{fig:fig8}C, a microgel with the same fraction of crosslinker ($c=0.05$) but surface charge distribution, in interaction with large NPs ($D=8\,\sigma$) is shown to display a rather homogeneous corona that is pushed to a rather compact structure by the attached NPs. The inset also shows that long corona chains tend to wrap around the NPs in order to maximize the contact with charged monomers. In addition, in this situation, counterions do not play a signifcant role, since they mostly remain free in suspension. Instead, a very different behavior is reported in Figure \ref{fig:fig8}D, illustrating a low-crosslinked microgel ($c=0.01$) with surface charge distribution interacting with small NPs ($D=2\,\sigma$). In this case, the shrinking is not so evident from the outside corona that remains rather heterogeneous, but it is interesting to see the large screening effect of the positive counterions, which tend to cover the NPs both close and outside the microgel.
Despite the variety of microgel--NPs assemblies, the resulting shrinking behavior is always well-described in terms of $Q_{complex}$ in all considered cases, suggesting the generality of our findings.

\section{Conclusions}
In summary, this work provides the first numerical characterization of realistic 
microgels in electrostatic interaction with oppositely charged NPs. The simulations are complemented by experiments performed on cationic microgels, that assemble with negatively charged gold NPs, of interest for plasmonic applications. 
The aim of our investigation is to establish a microscopic model able to describe the observed microgel deswelling upon NPs addition, either by their adsorption on the microgel corona or by incorporation within the inner regions of the polymer network. The present study provides a strong confirmation that, independently of the NPs arrangement, the microgels always tend to shrink within all studied conditions. These findings are consistent with previous experimental studies, performed for different mesh sizes of the microgels, \latin{e.g.}~by changing crosslinker concentration, or for different NPs charge or size. 
In the present work, we vary all these control parameters and we also tune the arrangement of the NPs by changing the location of the ionic groups within the microgel, thus providing a unifying framework for all the different observations. Indeed, we find a general behavior where the microgels deswell up to a minimum size, which corresponds to a given amount of added NPs, varying with the specific characteristics of the system, after which a plateau behavior occurs. Above this limit, most NPs remain free in suspension because the microgel--NPs complex has now reached an overall neutrality, and long-range electrostatic interactions become repulsive when additional NPs are incorporated in the microgel.

While these are well-established observations, we here provide a description of the deswelling in terms of the microscopic changes taking place within the microgel structure. In particular, we find that when NPs are mainly located on the surface, the shrinking occurs by the folding of the more external chains and the reduction of the microgel radius of gyration is well-reproduced by the average shrinking of the gyration radii of individual chains. This is in accordance with the hypothesis that adsorbed NPs shield the repulsion between charged monomers, analogously to an increase of the ionic strength \cite{hain2008,das2007,davies2010,sennato2021}. Instead, when NPs penetrate inside the microgel, the individual rearrangements are not sufficient to describe the observed deswelling, but rather a collective reorganization of the chains take place, that is driven by the connectivity of the network. So, in this case, chains on average migrate towards the microgel center of mass, mainly because of their electrostatic interactions with the inner NPs. This second scenario is more consistent with the proposed role of NPs acting as additional crosslinkers for the polymeric network\cite{gawlitza2013,pich2005,hou2014,hain2008,das2007,bradley2011}. 
In both cases, however, the shrinking process can be explained in terms of the additional pressure acting on the microgel, that takes place upon addition of the NPs.
Such a pressure reaches a maximum and then decreases when the interaction of the overall complex with non-adsorbed NPs becomes repulsive upon further addition of NPs.

Finally all the present simulation data can be rescaled onto a single master curve, when we consider the charge of the microgel--NPs complex as the variable of interest. This plot highlights that the deswelling process roughly stops when $Q_{complex}$ reaches zero. Interestingly, the variation of the characteristics of microgel and NPs gives rise to very different morphologies for the resulting complexes, being these more or less compact or heterogeneous, undergoing more or less screening by the counterions, and most of all having NPs arranged differently within the network. Nonetheless, all these different cases are found to follow the same master behavior, that is in very good agreement with experiments, suggesting a way to control the deswelling properties of these nano-complexes by means of their microscopic features. 
The emergence of such a unified behavior further points towards a novel criterion to determine the amount of adsorbed NPs based on the evaluation of the effective charge of the complexes, with no need for their direct visualization: we have indeed shown that the number $N_{ads}$ of adsorbed NPs can be directly inferred when $Q_{complex}/Q_{microgel}>0$, since $N_{ads}=n$. Furthermore, the isoelectric point marks the deviation from complete adsorption pointing towards a very general criterion for maximizing the adsorbing efficiency of microgels.

These results are relevant for all those applications, from drug delivery to pollutants removal to catalysis, where the knowledge of the precise amount of the adsorbant is crucial. For instance, the fine control on the adsorption process would boost the accuracy of microgel-mediated patterning of plasmonic NPs, where the occurrence of defects, due to scarce incorporation in the microgel, interferes with the photonic properties of the array \cite{kumar2007}.
The practical translation of our criterion can drive the rational design of microgel--NPs assemblies by indicating a ``recipe'' to choose the proportions of the components, their charge, size, and morphology, with the advantage to reduce all those time-consuming development steps for the optimization of the system and removal of non-adsorbed NPs.

Of course, such nanostructured systems are mostly investigated in the literature for the possibility of modulating the  plasmonic properties of metallic NPs provided by the microgels responsivity. However, to be able to fully exploit these intriguing properties, we need to obtain a precise knowledge of the location of the NPs within the network, and to control this, \latin{e.g.}~in response to temperature. This would allow the fine tuning of the distance between adsorbed NPs and in turn the coupling of their plasmon resonances, of great interest for sensing applications. We thus plan to extend the present study to different temperatures across the Volume Phase Transition in the near future, to optimize the plasmonic response of the nanocomplexes.

In addition, while the microscopic mechanisms taking place may be generic, several other cases other than electrostatically-driven complexation have been reported~\cite{gawlitza2013,pich2005,pich2004}, and it will be interesting to address them in order to discriminate whether other qualitative situations may occur due to the interplay between attractive and repulsive interactions acting at different length scales.
However, these may be  in some cases related to specific chemical features of the complexes, and thus, it may be difficult to tackle them within a simple coarse-grained model, such as the present one. Instead, a case that would be interesting to study with the present framework is that of a much weaker electrostatic interaction, that could lead to reversible assembly and disassembly of NPs from the microgels. In the present simulations, the interactions are quite strong, even for the case $q=-10\,e$, so that reversible association occurs very rarely during the course of the simulations. However, we repeated simulations for independent runs and such non-equilibrium effects were found not to qualitatively affect the reported results. Nonetheless, we expect that reversible association could change the present behavior at least in some aspects and, therefore, it will be the subject of future investigations.

Finally, the present numerical study deals with just a single microgel, but it will be important in the future to extend it to include at least a few microgels to see whether NPs-bridging and cluster formation could take place under specific conditions. This could provide another way for the fine tuning, on one hand, of the NPs-assembly to optimize their plasmon coupling, and, on the other hand of the packing of microgel--NPs complexes in ordered arrays.
Our study therefore is meant to represent the first crucial step laying the foundation for a more systematic investigation, where microgel--NPs nanostructured complexes will be designed and optimized for applications.

\section{Materials and Methods}

\subsection{Sample preparation}
Our experiments are performed on two different types of cationic microgels, the first with molar fraction of crosslinker monomer $c=0.05$ and molar fraction of charged monomer $f=0.032$, the second with $c=0.01$ and $f=0.02$. For the synthesis, we use the surfactant-free radical polymerization previously detailed \cite{truzzolillo2018, sennato2021}.
Briefly, we dissolve 236.67 mg of NIPAM monomers (Sigma-Aldrich, $\text{MW}=113.16$ Da) and the crosslinker N,N'-methylene-bis-acrylamide (BIS, Sigma-Aldrich, $\text{MW}=154.17$ Da) in 26.5 ml of deionized water. Separately, the ionic initiator 2,2'-Azobis(2-methylpropionamidine) dihydrochloride (AIBA, Sigma-Aldrich, $\text{MW}=271.19$ Da) is dissolved in 1.2 ml of water.
The solution containing NIPAM and BIS is bubbled with argon for 30 minutes and, after heating up to 70 °C, the initiator solution is added. 
In this way, in the final solution, the mass fraction of NIPAM is 0.0085.
At 70 °C, AIBA undergoes homolytic cleavage forming two radicals. Each of them reacts with a NIPAM monomer and produces a new radical, giving rise to the polymerization reaction.
Therefore, after starting the reaction, AIBA initiator remains attached to the backbone of the microgels and provides them positive charge, due to the protonation of amine groups.
After 6--hour reaction, the obtained dispersion is cooled down to room temperature and filtered through glass wool. 
To prevent bacterial growth, \ce{NaN3} (Sigma-Aldrich, $\text{MW}=65.01$ Da) is added to the concentration of 2 mM.
The final volume fraction $\varphi$ of the microgels in the dispersion is determined by viscosimetry measurements at 25 °C following the method described in Ref.~\citenum{truzzolillo2015}. We obtained $\varphi = 1.4 \pm 0.1 \times 10^{-1}$ for the sample with $c=0.05$, and $\varphi = 1.6 \pm 0.1 \times 10^{-1}$ for the sample with $c=0.01$. The number density of the microgels is estimated by $n_{mg} = \varphi/v_{mg}$, where $v_{mg} = \frac{4}{3}\pi R_{H}^3$ and $R_H$ is the hydrodynamic radius of the microgels measured by DLS at the same temperature ($R_H = 343\pm 9$ nm for $c=0.05$ and $R_H = 410\pm 7$ nm for $c=0.01$). We obtain $n_{mg} = 8.28 \times 10^{11}$ mL$^{-1}$ for $c=0.05$ and $n_{mg} = 5.54 \times 10^{11}$ mL$^{-1}$ for $c=0.01$.

To prepare microgel--NPs samples, we use spherical gold NPs with nominal diameter of 20 nm (Ted Pella). Gold NPs are stabilized by a citrate capping that provides them with a negative charge. The nominal number density of NPs is $n_{NP}=7.0\times10^{11}$ mL$^{-1}$. Such a low concentration guarantees that possible effects on the ionic strength of the final samples due to ions release from the NPs surface are negligible.
We dilute separately the microgel dispersion 250 times in 0.4 mM \ce{NaN3} and the NPs one in MilliQ water to obtain the desired number density. The suspensions are then stored at room temperature to avoid effects of thermal gradients on the interaction of the microgels with NPs during the following step.
Subsequently, we appropriately mix the two components and gently agitate the solution by hand, to obtain the final samples, where the microgels--NPs number ratio $n = n_{NP}/n_{mg}$ varies in the range 1 -- 200. The concentration of \ce{NaN3} in the prepared samples is 0.2 mM, low enough to exclude any effect of the
ionic strength on the microgel swelling.

\vspace{3cm}
\subsection{Experiments}

The morphology of the samples is studied by transmission electron microscopy, using a Tecnai G$^2$ 12 TWIN (FEI Company) that operates at 120 kV, equipped with an electron energy loss filter (Biofilter, Gatan Inc.) and a slow-scan charge-coupled device camera (794 IF, Gatan Inc.).
For imaging, 20 \textmu l of each sample is deposited at room temperature on a 300--mesh copper grid covered by a thin amorphous carbon film. When useful for better visualization, samples are stained by phosphotungstic acid, by adding 10 \textmu l of 2\% aqueous solution (with pH adjusted to 7.3 using 1 N \ce{NaOH}) to each deposition.
To evaluate the number of NPs adsorbed to each microgel, we count the number of dark spots (NPs) placed in correspondence of the opaque area individuating the microgel. For each sample, we average this quantity over at least 20 microgels.

The hydrodynamic radius $R_H$ is measured at 25 °C by DLS, employing a NanoZetaSizer apparatus (Malvern Instruments LTD) equipped with a He-Ne laser (5 mW power, 633 nm wavelength), that collects light at an angle of 173°.
The acquired intensity autocorrelation functions are analysed by means of the NNLS algorithm \cite{lawson1974} to extrapolate decay times, which are used to determine the distribution of the diffusion coefficients $D$ of the particles. Diffusion coefficients are then converted in intensity-weighted distributions of $R_H$ using the Stokes--Einstein relationship $R_H = k_BT/6\pi\eta D$, where $k_BT$ is the thermal energy and $\eta$ the water viscosity.
The electrophoretic mobility $\mu_e$ is determined at 25 °C using the phase analysis light scattering method \cite{tscharnuter2001} of the same NanoZetaSizer apparatus, which is equipped with a laser Doppler electrophoresis technique.
Each value of $R_H$ and $\mu_e$ reported in this work is the average of a distribution obtained by at least 50 measurements. The associated error is the corresponding standard deviation. 

From $R_H$ and $\mu_e$ measurements, we compute the effective electrokinetic charge $q_\textit{eff}$ of NPs and of microgel--NPs complexes. For spherical particles and low ionic strength ($ak_D<<1$, where $a=R_H$ is the particles radius and $k_D^{-1}$ is the Debye screening length), as in the case of the gold NPs dispersed in MilliQ water (SI, subsection S1.3), the H\"uckel relation gives $q_\textit{eff}=6\pi\eta a\mu_e$, where $\eta$ is the viscosity of water \cite{hunter}.
For microgel--NPs samples, instead, $k_D^{-1}\simeq20$ nm due to the presence of 0.2 mM \ce{NaN3}, and therefore a reliable estimate of $q_\textit{eff}$ needs to account for the Henry's correction\cite{hunter}, that reads\cite{ohshima1994}:
\begin{equation}
\label{eq:henry}
q_\textit{eff} = 6\pi\eta\, a\mu_e\,\frac{1+ak_D}{f(ak_D)}
\quad \text{, } \quad
f(ak_D) =  1 + \frac{1}{2\left[ 1 + \frac{5}{2\,ak_D \left( 1+2\,e^{-ak_D} \right) } \right]^3}
\end{equation}

\subsection{Numerical simulations}
We use coarse-grained microgels consisting of fully-bonded, disordered polymer networks of $N_{mg}=14000$ spherical beads of diameter $\sigma$ and mass $m$, which set the length and mass units. Microgels are prepared by the protocol previously reported in Refs.~\citenum{gnan2017} and \citenum{ninarello2019}, that allow to faithfully reproduce the experimental structure of the network. All the beads, that represent polymer segments, interact via the well-established Kremer–Grest bead-spring model~\cite{grest1986}, and therefore experience a steric repulsion modelled by the Weeks–Chandler–Anderson (WCA) potential:
\begin{equation}
\label{eq:wca}
V_{\text{WCA}}(r_{ij})  =  
\begin{cases}
4\varepsilon\left[\left(\frac{\sigma_{ij}}{r_{ij}}\right)^{12}-\left(\frac{\sigma_{ij}}{r_{ij}}\right)^6\right]+\varepsilon & \quad \text{if} \; r_{ij} \le 2^{1/6}\sigma_{ij}  \\
0 & \quad \text{if} \; r_{ij} > 2^{1/6}\sigma_{ij}
\end{cases}
\end{equation}
where $r_{ij}$ is the center-to-center distance between a given pair of interacting particles, $\sigma_{ij}$ is the sum of the two radii ($\sigma_{ij} = \sigma$ for the beads composing the microgel), and $\varepsilon$ sets the energy scale.
Additionally, bonded particles interact via the Finitely Extensible Nonlinear Elastic (FENE) potential \cite{del2019,kremer1990}:
\begin{equation}
\label{eq:fene}
V_{\text{FENE}}(r_{ij})  = 
-\frac{1}{2}k_F{R_0}^2\ln\left[1-{\left(\frac{r_{ij}}{R_0}\right)}^2\right]\; , \quad r_{ij} < R_0 \end{equation}
where $R_0=1.5\,\sigma$ and $k_F=30\,\varepsilon/\sigma^2$ are the maximum extension and the spring constant of the bond, respectively. The bonds cannot break during the simulation, mimicking strong covalent binding. Beads that are linked via the FENE potential to two neighbours represent segments of NIPAM chains, while crosslinkers have fourfold valence. As for experiments, we analyse two values of the crosslinker concentration, $c = 0.01$ and $c = 0.05$. 

To mimic the ionic groups of AIBA monomers, we provide a fraction $f$ of the microgel beads with a positive charge. We use $f=0.016$ and $f=0.032$, to match the nominal charge content in experiments for microgels with $c = 0.01$ and $c = 0.05$, respectively. We use both random and surface charge distributions. In the former case, charged beads are randomly chosen throughout the network (except for crosslinkers), while for the latter case, we randomly choose charged beads only in the exterior corona of the microgel, i.e. where the distance from the microgel centre of mass is higher than $R_g$. The overall electro-neutrality is preserved by inserting an equivalent number of oppositely charged counterions, whose diameter is set to $\sigma_c=0.1\,\sigma$ to facilitate diffusion within the microgel network and to avoid spurious effects from excluded volume~\cite{del2019}.
Counterions interact among each other and with microgel beads through the WCA potential. Electrostatic interactions are given by the reduced Coulomb potential:
\begin{equation}
V_{\rm coul}(r_{ij})=\frac{q_i q_j \sigma}{e^{*2}\, r_{ij}}\,\varepsilon\; ,
\end{equation}
where $q_i$ and $q_j$ are the charges of the interacting beads ($+\,e^*$ for charged monomers of the microgel and $-\,e^*$ for counterions), $e^*=\sqrt{4\pi\epsilon_0\epsilon_r\sigma\,\varepsilon}$ is the reduced charge unit, $\epsilon_0$ and $\epsilon_r$ are the vacuum and relative dielectric constants. We adopt the particle-particle-particle-mesh method \cite{deserno1998} as a long-range solver for the Coulomb interactions.

NPs are represented by single negatively charged beads. We used two values of the charge, $q=-\,10\,e^*$ and $q=-\,35\,e^*$, and three diameters $D=2\,\sigma$, $D=4\,\sigma$ and $D=8\,\sigma$. Together with NPs, a corresponding number of negatively charged counterions ($q=-e^*$, $D=0.1\,\sigma$) is added to preserve the overall neutrality of the system. NPs interact among each other and with other beads through the WCA and Coulomb potentials.

Simulations are performed with the LAMMPS package~\cite{plimpton1995} at the temperature fixed by $k_BT=\varepsilon$.
The equations of motion are integrated with a time-step $\Delta t = 0.002\,\tau$, where $\tau = \sqrt{m\sigma^2/\varepsilon}$ is the reduced time unit. We use the Nos\'e-Hoover thermostat in the constant NVT ensemble for equilibration ($1000\,\tau$) and the Velocity-Verlet algorithm in the constant-energy ensemble for the production runs ($20000\,\tau$). The latter are used to extract the equilibrium averages of the observables of interest.

Microgels are characterized in terms of gyration $R_g$ and hydrodynamic $R_H$ radii. The gyration radius is calculated as
\begin{equation} \label{eq:Rg}
R_g=\left\langle\left[\frac{1}{N}\sum_{i=1}^N (\vec{r}_i-\vec{r}_{CM})^2\right]^{\frac{1}{2}} \right\rangle ,
\end{equation}
where $\vec{r}_i$ and $\vec{r}_{CM}$ are the positions of the $i$-th particle and of the microgel center of mass, and the sum is computed over all the $N$ monomers composing the microgel.
To compute $R_H$ from simulations, we use the ZENO software \cite{zeno}, that, based on an electrostatic-hydrodynamic analogy \cite{douglas1994}, derives the hydrodynamic properties of arbitrarily shaped objects using a probabilistic calculation by a Monte Carlo method.
We estimate $R_H$ of the microgel--NPs complexes by including in the calculation all the monomers of the microgel and the NPs bonded to charged monomers, identified as those with distance from the closest charged monomer lower than $1.25\,(D+\sigma)$.
Both the gyration radius and the hydrodynamic radius are used to evaluate the microgel shrinking at varying the number $n$ of NPs in the simulation, by the ratio to the same quantity computed on the bare microgel ($R_g^{(n)}/R_g^{(0)}$ and $R_H^{(n)}/R_H^{(0)}$). Based on $R_H$, we also define the swelling amount as $\Delta R_H/\Delta R_H^{(max)}=(R_H^{(n)}-R_H^{(min)})/(R_H^{(max)}-R_H^{(min)})$, where $R_H^{(max)}$ and $R_H^{(min)}$ are the maximum and minimum values of $R_H$.

For each particle type $j$ (e.g. monomers, ions, counterions or NPs), we compute radial density profiles as 
\begin{equation}\label{eq:profile}
\rho_j(r)= \left\langle \frac{1}{N_j}\sum_{i=1}^{N_j} \delta (|\vec{r}_{i}-\vec{r}_{CM}|-r) \right\rangle.
\end{equation} 
where $N_j$ is the number of particles of interest.

The total pressure of the system $P_{tot}$ and that acting on the microgel only $P_{mgel}$, are computed by the sum of the kinetic energy and the virial contributions:
\begin{equation}\label{eq:press}
P_j = \frac{N_jk_BT}{V_j}+\frac{1}{3V_j}\sum_{i=1}^{N_j}\vec{r}_i\cdot\vec{F}_i
\end{equation} 
where $N_j$ and $V_j$ are the number of particles and the volume of the system of interest (equal to $N$ and $V$ for the whole system and to $N_{mg}$ and $\frac{4}{3}\pi R_H^3$ for the microgel, respectively), and $\vec{F}_i$ is the total force acting on the $i$-th particle.

In order to study the microscopic mechanisms of the microgel-NPs interaction, we define chains as the set of consecutively bonded divalent monomers connecting two crosslinkers. 
The observables used to study each chain $k$ and its interaction with NPs are the number $\ell_k$ of monomers per chain. the total charge $q_k$, namely the number of charged monomers in the chain, the radial position $\vec{r}_k$ of the chain, defined as the distance of its center of mass from the one of the microgel, the distance $d_k$ from the closest NP, defined as the minimum monomer-NP distance among monomers of the chain, and the chain gyration radius $R_{gk}$, defined as in Eq.~\ref{eq:Rg}, where the sum is computed only over the monomers of the chain, thus $N$ and $\vec{r}_{CM}$ are replaced by $\ell_k$ and $\vec{r}_k$.\\
The charges of the microgel $Q_{microgel}$ and of the microgel--NPs complex $Q_{complex}$ are defined as the total charge of the microgel and of all the particles (counterions and NPs) embedded into it, identified as those with distance from the microgel center of mass shorter than $R_H$.

\begin{acknowledgement}
The authors thank Federico Bordi for fruitful scientific discussions and CINECA-ISCRA for HPC resources. D.T. and E.C. acknowledge financial support from the Agence Nationale de la Recherche (Grant ANR-20-CE06-0030-01; THELECTRA). S.S. and E.Z. acknowledge financial support from INAIL, project MicroMet (BRiC 2022, ID 16). 
\end{acknowledgement}

\begin{suppinfo}
Additional experiments: wide field electron microscopy images, electrophoretic mobility measurements, determination of NPs charge. Additional simulations data: density profiles of counterions, derivation and validation of Equation \ref{eq:rg_chains}, trends of $D_n$ for surface distribution and of $S_n$ for random distribution, deswelling master curve from gyration radii.
\end{suppinfo}

\bibliography{references}

\end{document}